# On invariance of specific mass increment in the case of non-equilibrium growth

L. M. Martyushev, A.P. Sergeev, P. S. Terentiev

Ural Federal University, 19 Mira Str., Ekaterinburg, 620002, Russia
Institute of Industrial Ecology, 20 S. Kovalevskoy Str., Ekaterinburg, 620219, Russia
leonidmartyushev@gmail.com

*Abstract*

It is the first time invariance of specific mass increments of crystalline structures that co-exist in the case of non-equilibrium growth is grounded using the maximum entropy production principle. Based on the hypothesis of the existence of a universal growth equation, with the use of dimensional analysis, an explicit form of the dependence of specific mass increment on time is proposed. Applicability of the obtained results for describing growth in animate nature is discussed.

*Introduction*

Interest in the study of pattern formation in the case of non-equilibrium growth (primarily, dendrite crystallization) originated long ago but still remains topical. First of all, this is connected with the fact that such processes determine many properties of melt solidification. In addition to applied interest, there are reasons of fundamental nature. So, scientists studying problems of the physics of non-equilibrium processes consider dendrite growth as an important example based on which an attempt to understand and investigate common regularities of dissipative non-equilibrium processes, which generate complex structures in both animate and inanimate nature, can be made. A dependence of mass on time for these growing structures represents one of the main characteristics of a non-equilibrium system.

The papers [1,2] present results of the experimental research on the dependence of mass on time during the growth of symmetrical and asymmetrical dendrites as well as the so-called seaweed structures in the case of non-equilibrium crystallization of $NH_4Cl$ from a supersaturated aqueous solution. Measurements were made both for the whole growing crystals and for their parts (for instance, separate dendrite branches). For crystals (and parts thereof) growing simultaneously and approximately at the same supersaturation (the so-called co-existing crystals), it was discovered that, in spite of a significant difference in mass $m$ and mass change rate $\dot{m}$, specific mass increments $\dot{m}/m$ are identical within the experimental error. It was also found that a specific mass increment can be well described using an empirical dependence of the form (the so-called DS model):

$$\dot{m}/m = a/t - b, \qquad (1)$$

where $t$ is time, and $a$ and $b$ are some coefficients. The coefficient $a$, according to Refs. [1,2], is rather universal and equals 1.8±0.1 for all the structures. The value of the coefficient $b$ depends on the supersaturation at which the growth takes place.

The above empirical results are very surprising and interesting but they lack theoretical discussion in their original papers. The objective hereof is to seek their explanation.

*Results and discussion*

Let us base our reasoning on the maximum entropy production principle that is presently well grounded and is commonly used for thermodynamic analysis of non-equilibrium systems (see, for example, the papers [3-7]). One of the principle's formulations applicable to the problems at hand is as follows [5-10]: in the case of non-equilibrium growth, two co-existing structures have the same specific entropy productions.

Using the classical expression of entropy production as a product of crystallizing matter's flux and a thermodynamic force (gradient of chemical potential) [3,5], the following can be written for the first and the second parts of a non-equilibrium growing crystal:

$$j_1 \Delta\mu_1 / R_1 = j_2 \Delta\mu_2 / R_2, \qquad (2)$$

where $j_1$ and $j_2$ are the fluxes of a crystallizing component moving towards the boundaries of the first and the second parts of the crystal, and $\Delta\mu_1 / R_1$ and $\Delta\mu_1 / R_1$ are the gradients of chemical potentials found at the distance of $R_1$ and $R_2$ (typical sizes of the first and the second parts of the crystal, respectively).

We can assume that the differences of chemical potentials in the vicinity of the co-existing parts at hand are approximately equal ($\Delta\mu_1 \approx \Delta\mu_2$). Using the law of mass conservation at crystal boundaries (accurate up to a constant factor), $j = \dot{m}/R^2$ (where $\dot{m}$ is the change of the crystal mass $m$ with time), the following shall be obtained:

$$\dot{m}_1 / R_1^3 \approx \dot{m}_2 / R_2^3 \qquad (3)$$

or, after dividing by crystal density:

$$\dot{m}_1 / m_1 \approx \dot{m}_2 / m_2. \qquad (4)$$

Thus, the application of the maximum entropy production principle allows obtaining the equality of specific mass increments for different co-existing parts of a non-equilibrium growing crystal, as is observed in the experiment.

Let us note that a dependence of the form (4) is typical not only in the case of non-equilibrium growing crystals but is also very common in animate nature. This is the so-called law of ontogenetic allometry stating the proportionality of relative changes in the masses of an organism's growing parts. G. Cuvier (1798) was at the origins of this law, whereas J. Huxley (1932) gave its final formulation for animate systems based on the analysis of empirical data [11]. Therefore, the above derivation of Eq.(4) based on common thermodynamic (rather than particular) statements is very valuable. On the other hand, the equation (1) with $b=0$ is also known in biology. This is the law of growth empirically obtained by Schmalhausen in 1927-1930-ies following the study of embryonic growth [12-16]. According to Schmalhausen, such a power formula provides much better results than the exponential formulae. The studies [15,16] show that (1) can also be applied in order to describe the change of mass with time in the case of post-embryonic growth. The results obtained in the above papers make it possible to formulate a hypothesis of the existence of a universal growth equation for various objects of animate and inanimate nature which describes the change of mass with time. Let us try to analytically derive this equation on the basis of this hypothesis.

A system developing in time is extremely complex from morphological and kinetic perspective. To consider such complex and poorly formalized systems, dimensional analysis is traditionally used [17]. Using this analysis, fundamental relations between quantities that determine the process can be found correctly. Let the change of some mass with time $\dot{m}$ be the main sought-for quantity in the growth equation. First of all, this quantity must be dependent on the mass $m$ of a growing object (crystal, organism, etc.): the bigger the mass, the more it changes (in its absolute value) during the growth. There are other quantities that, alongside with mass, define $\dot{m}$. Let us use $\Omega$ to designate a set of these quantities. Then let us remind that the we have accepted the hypothesis that the law of growth is universal, i.e. growth equations for various objects (crystals, animals, plants, etc.) and for objects of other organizational levels (crystalline parts, plant organs, etc.) must have identical forms. Thus, no dimensional constants or variables specific for a particular growing object shall be present in such a universal equation. For example, frequency of incident radiation and the Planck constant could be important dimensional quantities for plants, coefficients of diffusion in a medium and kinetic coefficient of crystallization could be important for crystals, etc. For the growth of one system, some dimensional quantities can be essential, whereas, for the growth of other systems, they can have only secondary meaning or no meaning at all. As a result, based on the hypothesis of universality, almost all quantities that could determine growth rate in every specific case have to be excluded from consideration. However, there must be a quantity(ies) with the dimension of time in $\Omega$ as follows from the considerations of dimensionality. It seems that time of growth (age, if the growth of animate systems is considered) is the essential quantity to select. This quantity can be introduced for any growing object; and the influence of $t$ on $\dot{m}$ is obvious[1]. A universal equation of growth, therefore, certainly contains age as the sought-for dimensional variable. Other quantities with the dimension of time can describe a process of changing properties in the medium of growth. Firstly, let us consider the most general case when medium properties are invariable in time. This approximation evidently hold true for the very initial stage of any growth. Consequently, for this case, we draw a conclusion that $\dot{m}$ is a function of $m$ and $t$ only and has the following form (pursuant to the dimensional analysis):

$$\dot{m} = a \cdot m/t, \qquad (5)$$

where $a$ is some dimensionless constant.

---

[1] So, for instance, a crystal growing in a solution accumulates impurities and other defects on its surface with time, which significantly influence the rate of growth.

Thus, the growth equation (5) is obtained on the basis of the dimensional analysis following two assumptions: 1) an equation of growth is universal for various objects (of both animate and inanimate nature); 2) a change in a medium can be neglected.

The Eq.(5) can be easily specified for the case of changing growth conditions. We can make the simplest assumption that changes in a medium occur due to the growth itself and they are proportional to the mass of a growing body only. Hence, (5) can be rewritten as follows:

$$\dot{m} = a \cdot m/t - b \cdot m, \qquad (6)$$

where $b$ is some positive constant having the dimension of reverse time. This dependence conforms to the *DS* model (1) that was empirically proposed earlier.

Thus, using the maximum entropy production principle, the hypothesis of the existence of a universal growth equation, and the dimensional analysis, the present paper grounds invariance of specific mass increments for growing co-existing structures and obtains its explicit form, which was empirically obtained many times for growing objects of animate and inanimate nature.